\documentclass[twocolumn]{aastex631}

\newcommand{\Teff}{T$_{\rm{eff}}$}
\usepackage{amssymb,gensymb,times,graphicx,morefloats,color}
\usepackage{amsmath}
\usepackage{graphics,graphicx,url,verbatim,xspace}

\usepackage{changepage}
\usepackage{mathtools} 
\usepackage{comment}
\usepackage{gensymb}
\usepackage{hyperref}

\newcommand\changed[1]{#1}

\usepackage{listings}
\usepackage{xcolor}
\definecolor{codegreen}{rgb}{0,0.6,0}
\definecolor{codegray}{rgb}{0.5,0.5,0.5}
\definecolor{codepurple}{rgb}{0.58,0,0.82}
\definecolor{backcolour}{rgb}{0.95,0.95,0.92}
\lstdefinestyle{mystyle}{
  backgroundcolor=\color{backcolour},   commentstyle=\color{codegreen},
  keywordstyle=\color{magenta},
  numberstyle=\tiny\color{codegray},
  stringstyle=\color{codepurple},
  basicstyle=\ttfamily\footnotesize,
  breakatwhitespace=false,         
  breaklines=true,                 
  captionpos=b,                    
  keepspaces=true,                 
  numbers=left,                    
  numbersep=5pt,                  
  showspaces=false,                
  showstringspaces=false,
  showtabs=false,                  
  tabsize=2
}
\begin{document}

\title{Direct Detection of Known Exoplanets in Reflected Light: Predicting Sky Position with Literature Orbit Solutions}

\correspondingauthor{Logan A. Pearce}
\email{lapearce@umich.edu}

\author[0000-0003-3904-7378]{Logan A. Pearce}
\affil{Department of Astronomy, University of Michigan, Ann Arbor, MI 48109, USA}

\author[0000-0002-2346-3441]{Jared R. Males}
\affil{Steward Observatory, University of Arizona, Tucson, AZ 85721, USA}

\author[0000-0002-9521-9798]{Mary Anne Limbach}
\affil{Department of Astronomy, University of Michigan, Ann Arbor, MI 48109, USA}

\begin{abstract}
The next generation of ground- and space-based observatories will enable direct imaging and characterization of cold, mature planets through thermal emission and, for the first time, reflected light detection. Known RV and astrometrically detected planets provide a known population for detection and characterization observations. However, many of the most promising targets lack orbital parameters of sufficient precision to confidently predict their location on relative to the star for a direct imaging campaign. We have developed \texttt{projecc}, an open source Python package designed to generate sky-plane planet location posteriors from literature orbit solutions. This tool aims to facilitate community preparation for direct imaging observations of known planets. In this work we describe \texttt{projecc} and use it to examine two case study systems relevant to reflected light imaging with ELTs: GJ~876~b, which we find has a well-constrained prediction, and Proxima Centauri b, whose location remains highly uncertain.%, as well as one potential target for \textsl{Roman} CGI, HD~219134~h, which we estimate has a 40\% probability of being in a detectable sky location at any given time. 
We provide a web app for exploring reflected light planet targets and their orbit solutions, including predictions from literature for 17 additional planets, located at \url{https://reflected-light-planets.streamlit.app/}. We also discuss future upgrades to \texttt{projecc}.

%We conclude $\sim20\%$ of Jupiter-like and $\sim85\%$ of small planets we examined need improved radial velocity or astrometric measurements to facilitate the most efficient direct imaging campaigns. %Finally, we provide recommendations to improve consistency and standardization of required details when publishing planet orbital parameters as we enter an era of integrating radial velocity, astrometry, and direct imaging techniques.
\end{abstract}


\section{Introduction} \label{sec:intro}

The upcoming generation of observatories has the potential to directly image cold and mature exoplanets for the first time. This was recently accomplished with JWST in the mid-infrared with the direct imaging of the emission from mature giant planet Eps Indi Ab \citep{Matthews2024epsIndiAb}. Mid-infrared observations with JWST have the potential to detect planets as cold and old as Saturn and Jupiter \citep[e.g. ][]{limbachNewMethodFinding2022, 2024jwst.prop.6122B, Limbach2024MEOW-WD}. Mid-infrared detection of mature planets may also be possible with ground-based observatories such as the LBT \citep{Wagner2021ImLowMassPlanets} and the next generation of ground-based ELTs (e.g. ELT-METIS, \citealt{Brandl2010ELT-METIS, Bowens2021ExoplanetsWithMetis})

In addition to directly imaging mature exoplanet via emission in the mid-infrared, both ground- and space-based platforms will pursue characterization of exoplanet atmospheres using reflected light for the first time. Space based facilities such as \textsl{Nancy Grace Roman} Coronagraph Instrument \citep[CGI,][]{Spergel2013WFIRST} and the planned Habitable Worlds Observatory (HWO) will achieve high contrasts ($\mathcal{O}\,10^{-7}-10^{-10}$), enabling the direct imaging of gas giants \citep[Roman-CGI,][]{Carrion-Gonzalez2021RomanReflLightTargets} and terrestrial planets \citep[HWO, ][]{Astro2020Decadal} analogous to those in our solar system. The large apertures, and correspondingly much better inner working angle than space based platforms, will also give the next generation of ground-based ELTs access to terrestrial planets via extreme adaptive optics (ExAO) and coronagraphic instruments (ELT: ELT-PCS, \citealt{Kasper2021-ELTPCS}; GMT: GMagAO-X, \citealt{Males2024GMagAOX}; TMT: PSI, \citealt{Jensen-Clem2021-TMT-PSI}). 

%Uninformed reflected light exoplanet searches with these facilities are highly time-consuming and expensive. %\todo{(I think there's some estimates of how long this will take with HWO? maybe cite that paper here)}. Although lower achievable raw contrast,  
Reflected light campaign efficiency can be significantly improved by imaging known planets detected \changed{via indirect methods (e.g. \textsl{Hipparcos--Gaia} proper motion anomaly: AF~Lep~b, \citealt{Franson2023-AFLepb, DeRosa2023-AFLepb, Mesa2023-AFLepb},  HIP~99770~b, \citealt{Currie2023-HIP99770b}; radial velocity: $\epsilon$~Indi~A~b, \citealt{Matthews2024epsIndiAb}, $\beta$~Pic~c, \citealt{Nowak2020BetaPicc-direct-detection}, HD~206893~c, \citealt{Hinkley2023-HD206893c}; disk morphology: $\beta$~Pic~b, \citealt{Lagrange2010-BetaPic-b-direct-discovery}, TWA-7~b, \citealt{Lagrange2025-TWA7b}).} Observing efficiency is further improved if the planet's location is known beforehand \citep[e.g.][]{Carrion-Gonzalez2021RomanReflLightTargets, Carrion-Gonzalez2021-ConstrainingEpsErib}. 
Astrometric and RV detected planets with estimates of (some) orbital parameters and (minimum) mass enable estimating radius, phase, and prediction of their location on-sky at a given epoch, facilitating the most efficient observing campaign planning.

Due to the high contrasts involved (star-planet contrasts $\mathcal{O} 10^{6}$ -- $10^{9}$), long integration times and advanced wavefront control methods are required. Many speckle suppression techniques operate in a specific focal plane region that can correspond to a small range of orbital separations and position angles \citep{Kenworthy2025HCI-Review}. \changed{For example, the D-shaped ``dark hole" speckle suppression region formed by Electric Field Conjugation (EFC, \citealt{GiveOn2011EFC}) covers only a portion of the image plane over a range of separations. With a reliable prediction of the planet's focal plane position over time, this dark hole can be positioned to place the planet within the high contrast region. Without such a prediction, observation times must be increased by a factor of two or more to ensure the planet falls within the dark hole during some observations.} Additionally, orbital motion of shorter period planets, such as habitable zone targets, could impact efficiency. Accounting for orbital motion during observations and analysis boosts sensitivity  \citep{malesDirectImagingHabitable2013, LeCoroller2020KStacker}, and a confident constraint of orbital motion during and observation will facilitate confident detection.
Finally, precise knowledge of the planet's orbit is required to ensure we are attempting to observe it when it has the highest probability of detection. Knowing when the planet is at maximum elongation from the star and small scattering phase angle is essential for the best chance of detection.

%Several factors involved in planning for direct imaging observations of known planets require stronger constraints on a planet's orbital parameters than just the single-epoch star-planet separation. With the high contrasts involved, observations will need 10s--100s of hours \citep[e.g. ][]{Hardegree-Ullman2025BioverseO2-ELTs} to achieve adequate signal-to-noise for a confident detection, during which orbital motion of shorter period planets, such as habitable zone targets, could impact sensitivity \citep{malesDirectImagingHabitable2013}. Additionally, many speckle suppression techniques operate in a specific focal plane region \citep[e.g.][]{Haffert2023iEFC} that can correspond to a small range of orbital separations and position angles, requiring a confident prediction of the location of the planet in the sky plane at any given time to ensure it falls within the dark hole region.  Finally, precise knowledge of the planet's orbit is required to ensure we are attempting to observe it when it has the highest probability of detection. Knowing when the planet is at maximum elongation from the star and small scattering phase angle is essential to give the best chance of detection.

All of this means that when we attempt to directly image exoplanets discovered indirectly in either emission or reflected light, precise and accurate orbital element constraints facilitate the most efficient observation campaign on potentially highly oversubscribed ELTs and space-based platforms. Currently, many of the optimal first targets for reflected light campaigns in particular do not have published orbital elements of sufficient confidence for a robust prediction of the location of the planet over time. Continued dedicated radial velocity and astrometric campaigns to refine orbital parameters are of critical importance for the next generation of direct imaging with ground- and space-based platforms.

In this paper we present the new open source Python package \texttt{projecc}\footnote{\url{https://github.com/logan-pearce/projecc}\changed{DOI: \url{https://doi.org/10.5281/zenodo.15830011}}}, a tool for predicting the sky plane location of a planet at a given time for a set of orbital elements, incorporating measurement uncertainties. It is not an orbit fitter such as \texttt{orvara} \citep{Brandt2021orvara}, \texttt{RadVel} \citep{Fulton2018RadVel}, \texttt{orbitize!} \citep{blunt_orbitize_2020}, or \texttt{octofitter} \citep{thompsonOctofitterFastFlexible2023}, but a tool for visualizing published orbit solutions independent of the data source and fitting method. This allows us to determine (1) whether a planet's published orbital parameters can provide a confident prediction of location at a specific date and (2) how to optimize an imaging campaign based on its expected location or range of locations.  In Section \ref{sec:projecc} we describe the \texttt{projecc} package and define the orbital parameters and coordinate system used. 
In Section \ref{sec:examples} we use \texttt{projecc} to examine orbital solutions for two high priority ground-based reflected light targets, GJ~876~b and Proxima Centauri b, which illustrate the challenges of trying to directly image planets detected via RV and astrometry. Finally we end with recommendations for the future of planning reflected light imaging campaigns.

\section{Predicting Sky Position}\label{sec:projecc}

\subsection{Framework}
We developed the \texttt{projecc} Python package to work with orbit solutions, generally from indirect detections. It is designed for planning direct imaging campaigns, specifically predicting a planet's location in the image plane over time while accounting for orbital element uncertainties. We first define the orbital geometry and parameters used in our code, as these conventions vary yet are crucial for accurate planet location predictions.

\subsubsection{Definitions}\label{sec:projecc-definitions}

Orbital parameter reference frames and definitions can vary between (and even within) the RV, astrometry and direct imaging communities. Here we explicitly define our coordinate system and orbital parameter definitions and symbology, and how we employed published orbit solutions in our analysis.

We adopted the coordinate system of \citet{murray_keplerian_2010} (Figure 4). We established \textbf{($\hat x,\hat y, \hat z$)} denoting the orbit plane centered at the barycenter, where \textbf{$+\hat x$} defines the line from the barycenter to periastron (closest point of approach), \textbf{$+\hat y$} is perpendicular to \textbf{$+ \hat x$} in the orbit plane in a right-handed system, and \textbf{$+\hat z$} is orbit normal in the direction of the angular momentum vector. 

The orbit is described by seven parameters: semi-major axis, period, eccentricity, inclination, longitude of ascending node, argument of periastron, and time since a reference epoch. The semi-major axis ($a$) describes the size of the orbit and is half the distance from periastron to apastron (farthest point in the orbit plane). It is related to the period ($P$) and total mass ($\rm{M_{tot}}$) through Kepler's 3rd law: $\left(\frac{P}{1 yr}\right)^2  = \left(\frac{a}{1 au}\right)^3 \left(\frac{1 \rm{M}_{\odot}}{\rm{M_{tot}}}\right)$, where for a two-body system composed of one star (denoted with an asterisk) and one planet (denoted with subscript p) $\rm{M_{tot}} = M_* + M_p$. The position of a body at a given time is 
\begin{equation}
 \begin{pmatrix}
x\\
y\\
x\\
\end{pmatrix} =    
 \begin{pmatrix}
r \cos{f}\\
r \sin{f}\\
0\\
\end{pmatrix}
\end{equation}
where $r = \left(\frac{\rm{M}}{\rm{M_{tot}}}\right) \left(\frac{a (1-e^2)}{1+e\cos{f}}\right)$ with M representing either M$_*$ or M$_p$ depending on body of interest, $e$ is the eccentricity, and $f$ is the true anomaly, the angular location of the body relative to periastron. The true anomaly is computed as
\begin{equation}
    f = 2 \tan^{-1} \left(\sqrt{\frac{1+e}{1-e}} \tan(E/2) \right)
\end{equation}
where $E$ is the eccentricity anomaly, obtained from $M = E - e \sin E$, with $M$ as the mean anomaly, representing the angular location of the body in a circular orbit as a function of time. 

Projecting the orbit onto the sky plane, we established the coordinate system \textbf{($\hat X,\hat Y,\hat Z$)} where \textbf{$+\hat X$} is the reference direction and corresponds to the +declination direction, \textbf{$+\hat Y$} corresponds to the +right ascension direction, and \textbf{$+\hat Z$} forms a right-handed system with \textbf{$+\hat Z$} towards the observer. Note that this is the opposite of the common RV convention of \textbf{$+\hat Z$} being away from the observer. Transformation from the orbit plane to the sky plane uses the rotation matrices
\begin{equation}
    \bf{P}_x(\phi) =  \begin{pmatrix}
1 & 0 & 0\\
0 & \cos{\phi} & -\sin{\phi}\\
0 & \sin{\phi} & \cos{\phi}\\
\end{pmatrix}
\end{equation}
\begin{equation}
    \bf{P}_z(\phi) =  \begin{pmatrix}
\cos{\phi} & -\sin{\phi} & 0\\
\sin{\phi} & \cos{\phi} & 0\\
0 & 0 & 1\\
\end{pmatrix}
\end{equation}
Then:
\begin{equation}
 \begin{pmatrix}
X\\
Y\\
Z\\
\end{pmatrix} =    \bf{P}_z(\Omega)\bf{P}_x(i)\bf{P}_z(\omega)
 \begin{pmatrix}
x\\
y\\
z\\
\end{pmatrix}
\end{equation}
where $i$ is the orbit plane inclination relative to the sky plane with $i=90^\circ$ being edge on orbits, $i \in[0^\circ,90^\circ)$ corresponding to prograde (counterclockwise), and $i \in (90^\circ,180^\circ]$ corresponding to retrograde (clockwise); $\Omega$ is the longitude of ascending node, the angular location of the point where the orbit crosses the sky plane from \textbf{$-\hat Z$} to \textbf{$+\hat Z$} with $\Omega$ increasing counter clockwise from \textbf{$+\hat X$} (east of celestial north); $\omega$ is the argument of periastron, defining the angular location in the orbit plane from the ascending node to periastron. The final parameter is the timing parameter, which defines the body's position at a reference time from which the other orbit parameters determine its future location. Since the position of the body is set by the mean anomaly, which is the angular location since periastron, we use the epoch of periastron passage, $T_0$, as the reference time. At $T_0$, the body is at periastron, and $M = 0$. The mean anomaly is given by $M = 2\pi \frac{\Delta t}{P}$, where $\Delta t = t - T_0$ is the time since periastron passage, with $t$ being the time of interest and $P$ the orbital period. Table \ref{tab:orbital-elements} summarizes the orbital parameters and their units. 

% \begin{figure}
% \centering
% \includegraphics[width=0.48\textwidth]{MC-Fig4.png}
% \caption{\small{Reproduction of \citet{murray_keplerian_2010} Fig 4. showing the coordinate system adopted for our code. Note that \textbf{($\hat X,\hat Y,\hat Z$)} forms a right-handed system, which is the opposite of the common RV convention of \textbf{$+\hat Z$} being away from the observer.}}
% \label{fig:MCFig4}
% \end{figure}

\begin{table*}[]
    \centering
    \begin{tabular}{lccc}
       Parameter & Symbol & Unit & Prior \\
       \hline
        Semi-major axis & $a$ & au &  $\mathcal{N}(\mu,\sigma)$  \\
        Eccentricity & $e$ & -- & $\mathcal{N}(\mu,\sigma)$  \\
        Period & $P$ & days & $\mathcal{N}(\mu,\sigma)$  \\
        Inclination & $i$ & deg & $\cos(i)\; \mathcal{U}[0,\, 0.9845]$ or $\cos(i)\; \mathcal{U}[-1,\, 1]$  \\
        Argument of Periastron & $\omega_p = w_* + 180^\circ$ & deg & $\mathcal{N}(\mu,\sigma)$  \\ 
        Longitude of Nodes & $\Omega$ & deg & 0.0, $\mathcal{N}(\mu,\sigma)$, or $\mathcal{U}(0^\circ,360^\circ)$  \\
        Epoch of Periastron Passage & $T_0$ & BJD & $\mathcal{N}(\mu,\sigma)$   \\
        Star Mass & M$_{*}$ & M$_{\odot}$ & $\mathcal{N}(\mu,\sigma)$ \\
        Planet Mass or M$\sin(i)$ & M$_{p}$, M$_{p} \sin(i)$ & M$_{\oplus}$ & $\mathcal{N}(\mu,\sigma)$ \\
        % Star effective temperature & T$_{\rm{eff}}$ & K & -- \\
        % Planet equilibrium temperature & T$_{\rm{eq}}$ & K & -- \\
        Planet radius & R$_p$ & R$_{\oplus}$ & -- \\
        Parallax & $\varpi$ & mas  & $\mathcal{N}(\mu,\sigma)$\\
        \hline
    \end{tabular}
    \caption{Orbital and system parameter definitions used in this analysis and the prior used in simulations, where $\mathcal{N}(\mu,\sigma)$ refers to a Gaussian distribution using the values given in the relevant reference. We assume reported values for $\omega$ refer to the star and apply a +180 deg offset to obtain $\omega$ for the planet, as this distinction is often not specified in the literature.}
    \label{tab:orbital-elements}
\end{table*}

\subsubsection{Orbital Parameter Conventions and Potential Misinterpretations}
The above definitions are independent of whether they describe the star's or the planet's motion around the barycenter. For radial velocity and astrometric measurements, the star's motion around the barycenter is typically used to determine orbital parameters. In contrast, direct imaging relies on the planet's motion relative to the star, making the planet’s orbital parameters the primary focus. This difference in focus results in a few differences in coordinate definitions, which, if not carefully considered, can lead to errors. As noted above, our coordinate system defines \textbf{$+\hat Z$} in the opposite direction of the common convention in RV systems. Additionally, the argument of periastron for the planet is offset by 180$^\circ$ from the star, such that $\omega_p = \omega_* + 180^\circ$ \citep{Perryman2011ExoplanetHandbook}. In many cases it is unclear whether the reported $\omega$ refers to the star or the planet \citep{Brown2015TrueMasses, Xuan2020piMen}.Assuming the incorrect definition will impact the location of the planet in the sky plane and its viewing phase, and thus contrast, as a function of time, as the phase also depends on $\omega$ \citep[see][ for a detailed discussion]{Householder2022omega}. 
We assume $\omega = \omega_*$ where not specified, requiring a 180$^\circ$ offset when computing the orbit of the planet, following \citet{Carrion-Gonzalez2021RomanReflLightTargets}. 

Additionally, \citet{Carrion-Gonzalez2021RomanReflLightTargets} and \citet{Householder2022omega} both discuss how differences in definitions for the ascending node, \textbf{$+\hat Z$} convention, and the reference direction for $\omega$ can introduce systematic shifts in $\omega$. Unless stated in the relevant publication, we assume the RV convention where \textbf{$+\hat Z$} points away from the observer, and apply a factor of $-1$ where relevant. Additionally, we assume that $\omega$ is defined as presented here.
A standardized, clearly defined convention for orbital parameter notation and coordinate systems is crucial for accurately combining RV, astrometric, and direct imaging observations.

\begin{figure*}
\centering
\includegraphics[width=0.9\textwidth]{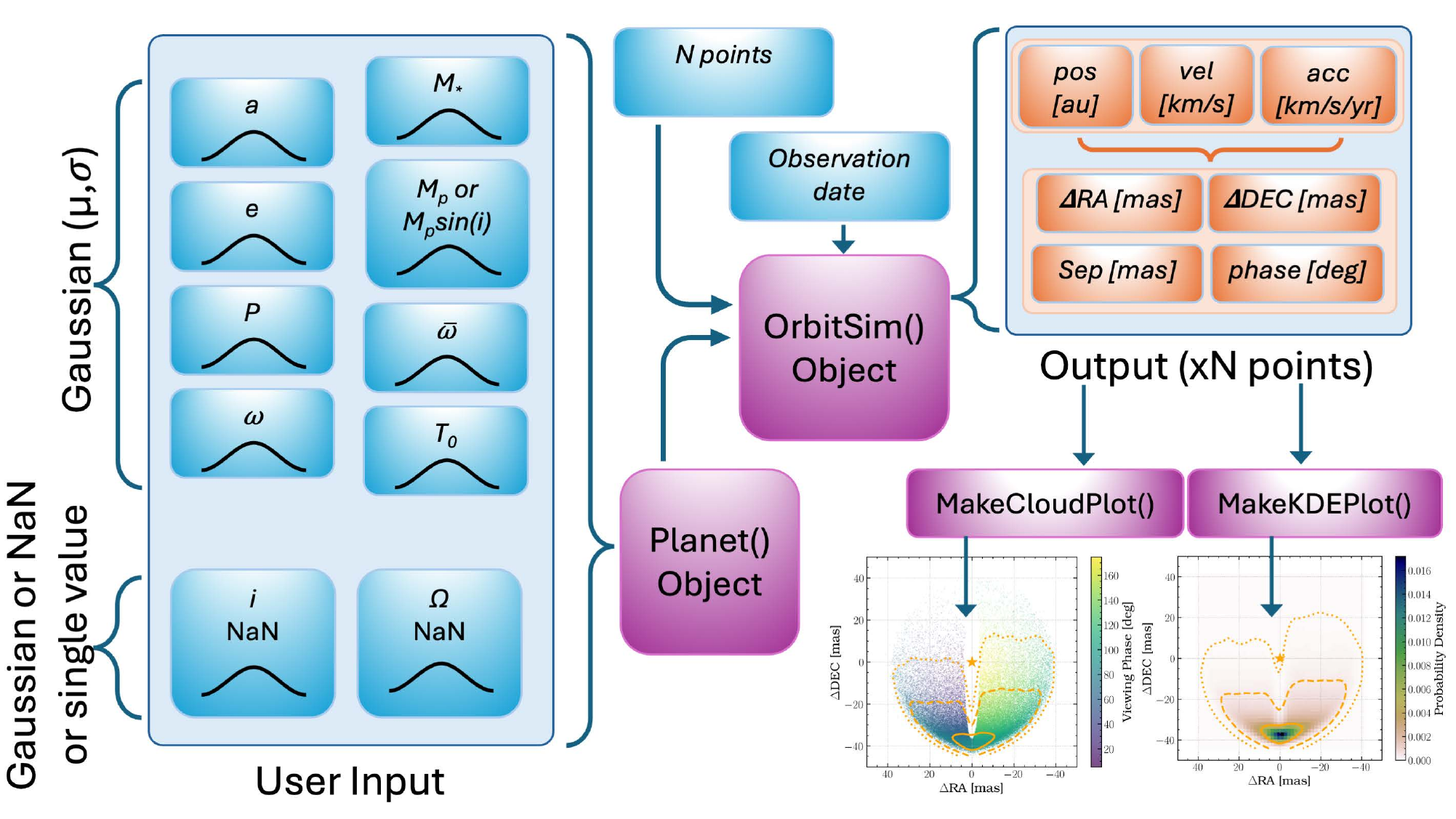}
\caption{\small{Schematic diagram of the function of the \texttt{projecc} package. Users supply orbital parameter inputs to the \texttt{Planet} class. Then users supply the \texttt{Planet} object, an observation time, and number of simulation points to the \texttt{OrbitSim} class, which produces a set of $N$ simulated realizations of the planet position at the observation date drawn from the orbital parameter distributions. The \texttt{OrbitSim} object contains $1xN$ vectors of sky plane position, separation, and phase which users can supply to the provided plotting tools.
}}
\label{fig:flowchart}
\end{figure*}

\subsection{\texttt{projecc}}

Now that the coordinate system is defined, we proceed to discuss \texttt{projecc}, the code we developed to predict a planet's sky location from orbital elements derived from RV, astrometry, and other sources. \texttt{projecc} is available on GitHub and via PyPI\footnote{\url{https://pypi.org/project/projecc/}}. It is designed to have a simple interface for supplying orbital parameters and producing a prediction of planet location in the sky plane a given time. It is not an orbit fitter, it does not interface with data; the source of the orbital parameter constraints is immaterial. Except where indicated (discussed below), we assume that reported orbital parameters have Gaussian uncertainties. \texttt{projecc} does not currently accept asymmetrical error distributions. 

Figure \ref{fig:flowchart} illustrates the basic structure of the code. Users supply inputs to a \texttt{Planet} class, then supply the \texttt{Planet} class, an observation date, and the number of simulation points desired to an \texttt{OrbitSim} class, which produces the distribution of predicted locations and phases of the planet at the specified time.

\subsubsection{Inputs}

Users supply tuples of Gaussian distributed parameters ($\mu,\sigma$) for $a$, $e$, $P$, $\omega_p$, M$_p$ or M$_p \sin(i)$, M$_*$, T$_0$, and $\varpi$. If applicable, users must convert $\omega_*$ to $\omega_p$. All parameters must be in the units listed in Table \ref{tab:orbital-elements}.  Users also supply $i$ and $\Omega$, which take flexible inputs:

\textit{Inclination:} Setting $i = NaN$ results in $i$ being drawn from a uniform distribution of\changed{ $\cos(i)\; \mathcal{U} [0.0,0.985]$ (10$^\circ$--90$^\circ$)} or $\cos(i)\; \mathcal{U} [-0.985,0.985]$ (10$^\circ$--170$^\circ$), specified by a keyword. We avoid face-on orbits (which produce no RV signal). Limiting $i\le90^\circ$ removes degenerate solutions from motion in the clockwise direction. Alternatively, specifying $i = [\mu,\sigma]$ draws $i$ from a normal distribution.

\textit{Longitude of Nodes:} Since $\Omega$ is undefined for RV-only solutions, supplying $\Omega = NaN$ will cause $\Omega$ to be drawn from a uniform distribution $\mathcal{U} [0^\circ,360^\circ]$. Alternatively, a single fixed value or Gaussian tuple of $\Omega$ can be supplied.

Inputs are stored in a \texttt{Planet} object.

\subsubsection{Orbit Simulation}

Once the orbital parameters are assigned to a \texttt{Planet} object, the user generates a set of simulated realizations of the planet's position at a specific time by supplying the \texttt{Planet} object, a desired time in BJD, and the desired number of simulation points $N$ (default $N=100000$) to the \texttt{OrbitSim} class. The simulated orbit points are created via a Monte Carlo sampling from the inputs which are then converted to points in the \textbf{($\hat X,\hat Y,\hat Z$)} coordinate system using the equations in Section \ref{sec:projecc-definitions}. These points are stored \changed{as an attribute of the OrbitSim object} as the $N$x3 vector \texttt{pos}, where \texttt{pos[i,0]} = \textbf{$\hat X$} = decl., \texttt{pos[i,1]} = \textbf{$\hat Y$} = R.A, and \texttt{pos[i,2]} = \textbf{$\hat Z$} in au for the $i$th point. Similarly the \texttt{vel} and \texttt{acc} $N$x3 vectors, \changed{also stored as OrbitSim attributes,} store the velocity in km~s$^{-1}$ and acceleration in km~s$^{-1}$~yr$^{-1}$ in the same orientation. We employed a cubic approximation solver for Kepler's equation following \citet{Mikkola1987} \citep[See ][Sec 3.1.3 for a comparison of solvers]{blunt_orbitize_2020}. The \texttt{pos} vector is converted to R.A. and decl. in milliarcseconds, and the viewing phase angle at each simulation point is computed. We used the parametrization of \citet{Madhusudhan2012AnalyticModels} to compute phase angle:

\begin{equation}\label{eqn:alpha}
    \alpha = \cos^{-1} \left[ \sin(i) \sin(f + \omega_p)\right]
\end{equation}
where $i$ is inclination, $f$ is the true anomaly, and $\omega_p$ is the planet's argument of periastron. Phase angles $\alpha <90^{\circ}$ correspond to gibbous phases (blue in Figure \ref{fig:targets}), $\alpha >90^{\circ}$ correspond to crescent phases (red), and $\alpha = 90^{\circ}$ is quadrature (white).
\texttt{projecc} returns the position, separation, and phase angle arrays of all $N$ points. 

Once the orbit point simulation is complete, \texttt{projecc} can produce a scatter plot of the simulated points in the sky plane colored by the viewing phase at each point and bounded by contours marking regions containing 68\%, 95\%, and 99.7\% of points (1, 2, and 3$\sigma$; see Figure \ref{fig:Proxcenb-clouds}). Alternatively, users can plot probability density maps that visualize the likelihood of the planet's location in the sky plane at the specified date. The probability density function (PDF) is computed using \texttt{scipy.stats.gaussian\_kde} 2D kernel density estimation function \citep{virtanen_scipy_2020}, then normalized such that the PDF sums to one.

We encourage users to explore the online tutorials\footnote{{\url{https://doi.org/10.5281/zenodo.15830011}}} for examples of how to use \texttt{projecc}. 

\subsection{\texttt{projecc} Interactive App}

In addition to the \texttt{projecc} installable package on GitHub and PyPI, we have provided an interactive webapp via Streamlit\footnote{\url{https://streamlit.io/}}, called the Ground-based Reflected Light Imaging Planner (GRIP), found at \url{http://getagrip.space/}. GRIP provides an interactive version of Figure \ref{fig:targets} and an interface to allow users to create \texttt{projecc} simulation plots without installing \texttt{projecc}. Additionally, the app contains a number of literature orbit solutions for planets of interest for direct imaging campaigns, allowing users to rapidly view and compare orbit solutions. Appendix \ref{appendixA} details the use of the app.

\subsection{Planned Future Upgrades}

Currently the code is configured to accept only Gaussian distributions for most orbital parameters and treats each parameter as independent. Summary statistics such as mean/median and standard deviation or credible intervals are a poor way to capture the behavior of orbit parameter statistical estimators: (1) asymmetric and multi-modal posteriors are common from orbit fits, (2) there is significant correlation among parameters, such as period/semi-major axis, semi-major axis/mass, inclination/mass, and eccentricity/inclination, and (3) posteriors for some parameters, such as eccentricity \citep{Hogg2010EccentrcityDistribution, ferrer-chavez_biases_2021} and inclination \citep{Pourbaix2001-Screening, Makarov2025GaiaInclBias} show significant bias, such that the mean and variance are not reflective of the true value convolved with uncertainties. A more thorough statistical study of sky plane location requires the full posterior likelihood function or samplings from the posterior distribution of each parameter \citep{Hogg2010EccentrcityDistribution}. Future upgrades will enable users to supply covariance matrices or posterior distributions. \changed{We plan to support outputs from popular orbit-fitting packages, analogous to how \texttt{octofitter} accepts \texttt{orbitize!} posteriors.} %However it is difficult to obtain these from published orbit fit results as typically only mean or median and sigma or credible intervals are reported.

\changed{As instrument designs evolve, we plan to add the capability to plot regions corresponding to dark hole configuration and to report the fraction of orbit realizations within the dark hole at a given time. For example, the \textsl{Roman} CGI hybrid Lyot coronagraph dark hole region will be a 360$^{\circ}$ annulus spanning $\approx$3--9~$\lambda/D$ \citep{Akeson2019Roman}, making it straightforward to compute the fraction of simulated points within the dark hole and identify targets with the highest detection probability. This functionality will be added to both the webapp and the installable package in the near future, along with support for other instrument dark hole shapes (e.g., the \textsl{Roman} shaped pupil coronagraph and EFC) and user-defined custom dark hole regions.
}

\changed{We are also adding functionality for users to compute the fraction of points within a dark hole across a date range, enabling planning for optimum observing windows for a given target.}

\subsection{Comparison to Other Orbit Visualization Tools}

\changed{The orbit fitting packages \texttt{orvara} \citep{Brandt2021orvara}, \texttt{RadVel} \citep{Fulton2018RadVel}, \texttt{orbitize!} \citep{blunt_orbitize_2020}, and \texttt{octofitter} \citep{thompsonOctofitterFastFlexible2023} all provide tools for visualizing orbit posteriors over time, including planet separation and position angle, orbit tracks in the sky plane, and radial velocity, with observational constraints overlaid. In contrast, \texttt{projecc} does not fit orbital parameters, but predicts the planet's R.A. and Dec relative to the star on a specified date using existing orbit solutions and uncertainties, a capability not typically emphasized in these packages (though \texttt{orvara} supports it for proper-motion anomaly fits). \texttt{projecc} is designed to generate location predictions independent of the source of the orbit constraints, offering a simple interface for rapid analysis and observation planning. The GRIP webapp is also unique in enabling exploration without requiring local installation.
}

\changed{Both the intent and design of \texttt{projecc} and the GRIP webapp are similar to the functionality of \url{whereistheplanet.com}, which predicts the R.A. and Dec offset, separation, and position angle and uncertainties of directly imaged planet and brown dwarf companions at a specified date from a published orbit solution \citep{Wang2021whereistheplanet}. The \url{whereistheplanet.com} site hosts a number of pre-loaded resolved companions to select, and displays the astrometric information on a chosen date and the reference to facilitate future observation planning of those planets. The GRIP webapp is complementary in that the targeted planet populations do not overlap -- resolved, directly imaged companions vs closer in, indirectly detected planets. 
}

\begin{figure*}
\centering
\includegraphics[width=0.95\textwidth]{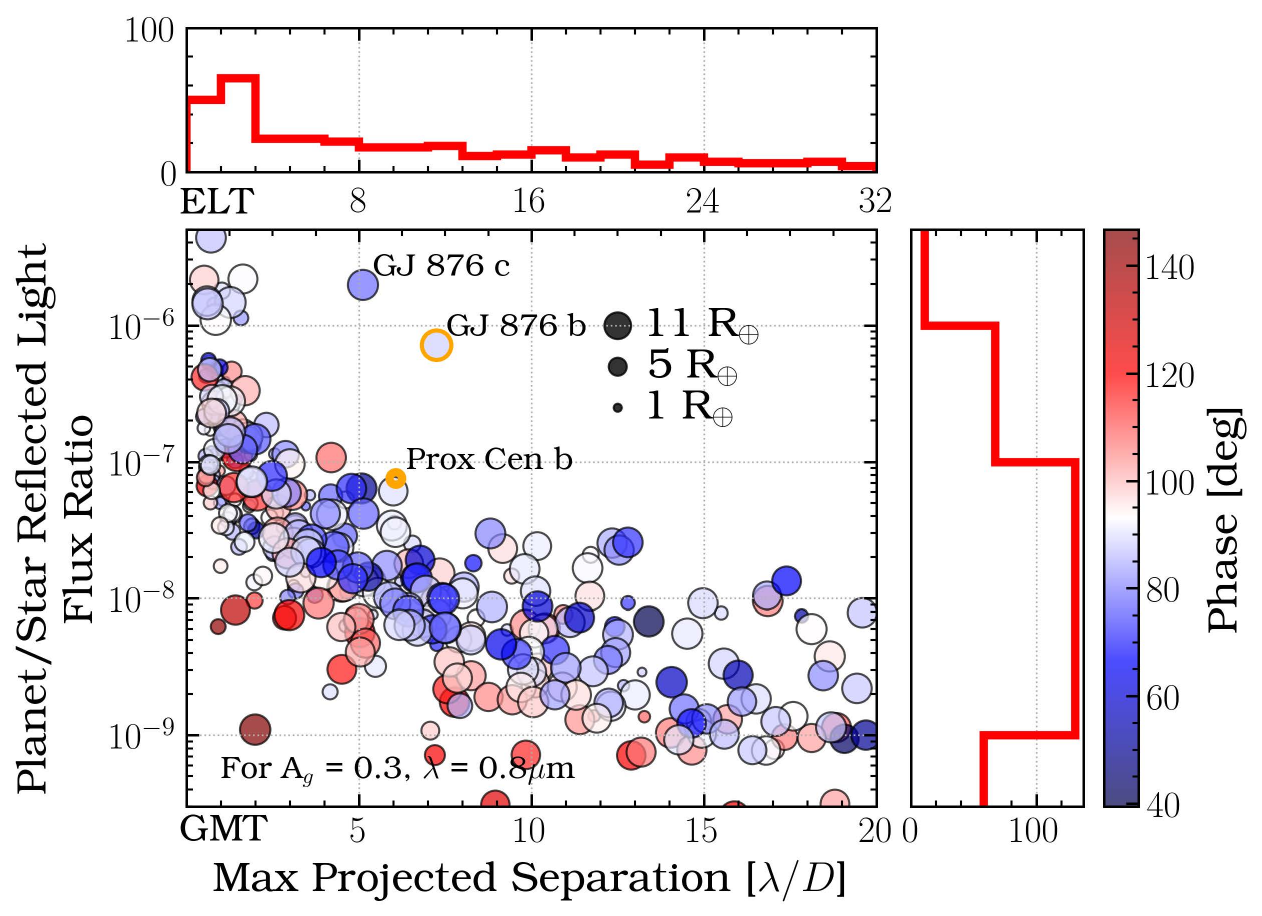}
\caption{\small{Approximately 400 of the nearest ($<$70~pc) RV detected exoplanets in the Exoplanet Archive (accessed Jan 2025) accessible in the southern hemisphere ($-65^{\circ}<$~dec~$<20^{\circ}$). The bottom x-axis shows the maximum planet-star separation in $\lambda$/D units for a GMT-sized primary (D$=25.4$~m) at $\lambda=800$~nm; the top x-axis shows the same for the ELT (D$=39$~m); the y-axis shows the Lambertian planet-star flux contrast using Eqn \ref{eqn: cahoy} assuming a geometric albedo A$_g=0.45$. %\footnote{  .  
Details for generating this list and plot are available at \url{https://reflected-light-planets.streamlit.app/} \changed{and in Appendix \ref{appendixReflLightTargets}}. The colormap shows the viewing phase at the planet's maximum elongation from the star.  The top histogram shows number of planets in each bin of size $\lambda/D$, while the right histogram represents the number of planets per bin spanning one order of magnitude in contrast. An interactive version of this plot is available at \url{https://reflected-light-planets.streamlit.app/}. The planets highlighted in this paper are marked with orange circles.
}}
\label{fig:targets}
\end{figure*}

\section{Case Studies in Estimating On-Sky Planet Positions using \texttt{projecc} 
}\label{sec:examples}

Predicting the on-sky location of an indirectly detected planet is challenging. The recent case of $\epsilon$~Indi~Ab is a striking example. \citet{Matthews2024epsIndiAb} directly detected $\epsilon$~Indi~Ab with \textsl{JWST} on the opposite side of the star and $\approx2\times$ further separation than predicted from the orbital solution that combined RV and astrometric signals \citep{Feng2019epsIndiAb, Feng2023epsIndiAb, Philipot2023UpdatedCahracterization}. They also derived a higher mass estimate. %It appears the discrepancy may be due to a too-short RV+astrometry baseline (24.8 years of observations vs. the planet's $\sim180$\,yr orbital period), with a longer observational baseline (29 years) giving a more consistent prediction \citep{Feng2024LessonsLearned}. 
This, being an early example of a direct detection campaign using an RV-derived orbit solution, highlights the need for robust orbit constraints.% and long observational baselines.
%it illustrates that we may not be well prepared to target known planets for reflected light campaigns.

\changed{We developed \texttt{projecc} as a tool to facilitate scrutiny of orbit solutions of future reflected light imaging targets and support efforts to refine these solutions prior to observing campaigns. Figure \ref{fig:targets} shows a set of RV detected planets comprising potential targets for ground based reflected light observations with upcoming ELTs.} In this section we examine literature orbits of two top priority systems for ground-based reflected light imaging, GJ 876 b and Proxima Centauri b, with \texttt{projecc} to illustrate challenges in predicting sky position, circled in orange in Figure \ref{fig:targets}. These targets were selected not only for their suitability for imaging campaigns but also to illustrate two extremes in predicting planet locations for direct imaging -- one case where the positions are well constrained from existing observations (GJ 876 b) and another where it remains highly uncertain (Prox Cen b). The Reflected Light Planets App provides the same plots for more than a dozen additional planets we examined in the course of this work -- 5 gas giants, 11 potentially terrestrial planets, and the three most accessible \textsl{Roman} CGI planets from \citet{Carrion-Gonzalez2021RomanReflLightTargets}, as well as in interactive version of Figure \ref{fig:targets}.

\subsection{The Gas Giants GJ 876 b and c}\label{sec:gj876}

The gas giant planets GJ 876 b and c are the two bright ($\approx1\times10^{-6}$) planets at $>5\lambda/D$ in Figure \ref{fig:targets}, making them the ideal first targets for reflected light campaigns. They orbit the moderately bright (m$_V \approx 10$) nearby ($\approx 4.7$~pc) early M star GJ 876 at 0.2 and 0.13~au respectively along with two other sub-Neptunes, d \citep{Rivera2005GJ876d} and e \citep{rivera2010}, not discussed in this manuscript. 

With RV and astrometry providing strong constraints on the planets' orbits, the system is among the best characterized.
Their close separations and Jovian masses (M$_b \sin i = 2.3$M$_{\rm{jup}}$, M$_c \sin i = 0.7$M$_{\rm{jup}}$) yield large RV signatures with several orbital parameter solutions in the Exoplanet Archive. GJ 876 b and c are believed to be coplanar with $i\approx50-60^{\circ}$, derived from enforcing \changed{dynamical stability} on multiplanet orbit fits \citep{Correia2010, rivera2010}. \citet{Benedict2002GJ876bMass} detected the astrometric wobble of GJ 876 due to GJ~876~b (the first astrometric detection of an exoplanet) using the \textsl{Hubble Space Telescope} Fine Guidance Sensor and derived a joint RV+astrometry orbit solution. GJ~876~b also has a non-single star (NSS) orbit solution \citep{Gaia2023-NSS} in \textsl{Gaia} DR3 \citep{GaiaDR3-2023}, providing an astrometric orbit fit from the \textsl{Gaia} NSS pipeline. It is listed as an \texttt{OrbitalTargetstSearchValidated} source, indicating that it was included on an input catalog for validating substellar NSS orbit solutions against literature and was found to be consistent with RV solutions. The \textsl{Gaia} DR3 NSS solution provides $P$, $T_0$, $e$, and Thiele-Innes constants, from which we computed the tabulated quantities using the \texttt{NSS Tools}\footnote{\url{https://www.cosmos.esa.int/web/gaia/dr3-nss-tools}} package \citep[][]{Halbwachs2023GaiaAstrBinary}.

\begin{figure}
    \centering
    \includegraphics[width=1\linewidth]{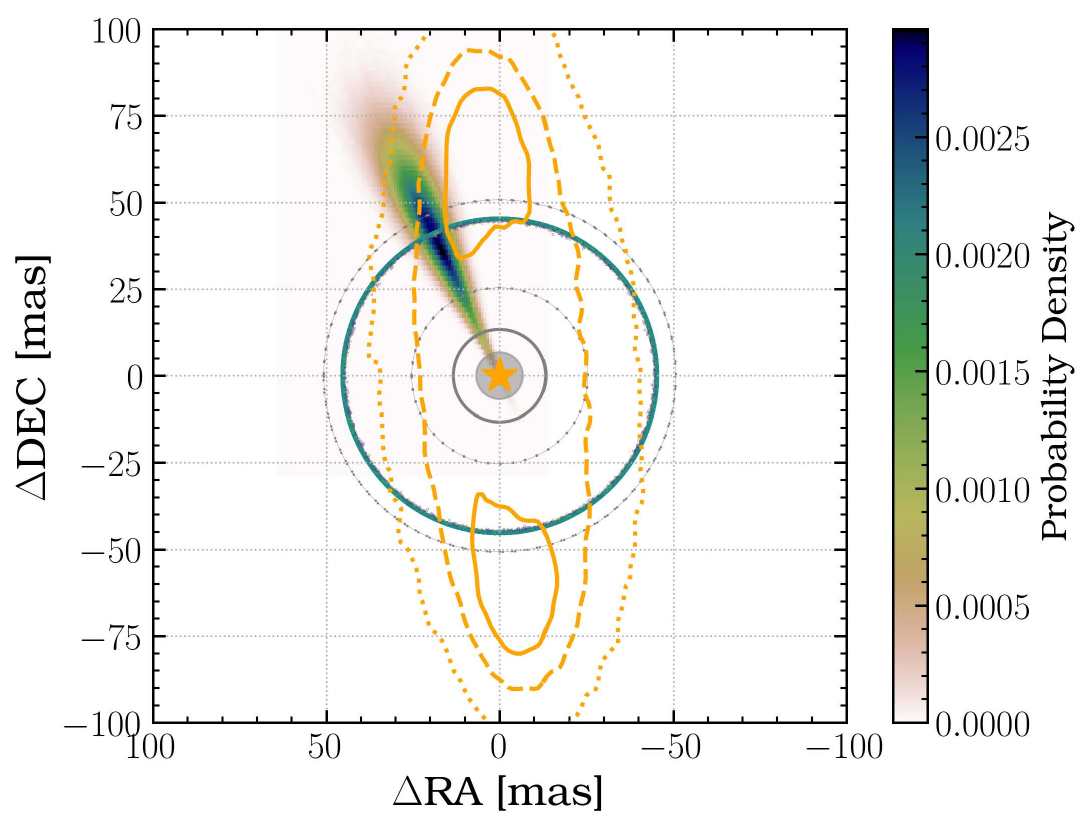}
    \caption{\texttt{projecc} predictions of the location of GJ~876~b on April 9th, 2025 for three orbital solutions, given in Table \ref{tab:gj876borb}. The RV solution of \citet{rivera2010} is shown by the scatter plot, which forms a ring of points, all viewed near quadrature, because $\Omega$ is not constrained and was drawn from a uniform prior on [0$^{\circ}$,360$^{\circ}$]. The blue to brown colormap shows the probability density of locations drawn from the \citet{Benedict2002GJ876bMass} RV+astrometry solution; the orange solid, dashed, and dotted curves show the 1, 2, and 3$\sigma$ contours respectively for the \textsl{Gaia} NSS astrometric solution. The solid grey circles show the size of 1 and 2$\lambda/D$ for an ELT-sized primary at $\lambda=$0.8$\mu$m, the dotted grey circles show 1 and 2$\lambda/D$ for a Magellan Clay Telescope primary ($D=6.5$~m) at  $\lambda=$0.8$\mu$m. The RV and RV+astrometry solutions are consistent while the NSS solution is only consistent at 2$\sigma$ and only on one side of the star.
    }
    \label{fig:gj876borbits}
\end{figure}

Table \ref{tab:gj876borb} shows orbital parameters for the Keck/HIRES RV-only \citet{rivera2010} fit (which is consistent with the other recent RV-only \citealt{Correia2010} solution), the RV+astrometry \citet{Benedict2002GJ876bMass} fit, and the astrometric-only NSS orbital solution. These three solutions are plotted using \texttt{projecc} simulation tools in Figure \ref{fig:gj876borbits}, predicting the planet location at a date when it is expected to be at maximum elongation from the star. The scatter plot ring shows the RV-only solution \citep{rivera2010}; the probability density shows the RV+astrometry solution \citep{Benedict2002GJ876bMass}, and the orange curves mark the 1-, 2-, and 3$\sigma$ contours from the NSS solution\footnote{In the absence of radial velocity information there is a degeneracy in $\Omega$ and $\omega$; the \textsl{Gaia} NSS orbit solutions limit $\Omega \in [0^\circ,180^\circ]$ if there is no spectroscopic solution \citep{Gaia2023-NSS}, as with GJ~876~b. Adjusting $\Omega$ by $\pm$180$^\circ$ does not meaningfully impact the probability contours here, so we tabulated the NSS solution of $\Omega = 4.2\pm4.5^\circ$, however $\Omega = 184.2\pm4.5^\circ$ is also a possible solution.}.

We see that the \citet{rivera2010} RV solution, with $\Omega$ unconstrained, tightly constrains the planet-star separation, and is consistent with the highest probability location from the \citet{Benedict2002GJ876bMass} solution, which constrains $\Omega$ but not separation. The NSS solution is consistent with both within 2$\sigma$, although without the additional orbit solutions the location of the planet would be poorly constrained with the NSS solution only. The NSS solution is derived by fitting a single Keplerian orbit \citep{Gaia2023-NSS}, and it is unclear if this solution accounts for perturbations from the three other previously known GJ~876 planets.  Their intersection provides a solid prediction of the location of GJ~876~b. 

\begin{table}[]
    \centering
    \begin{tabular}{cccc}
       Param& \citealt{Benedict2002GJ876bMass} & \citealt{rivera2010} &  \textsl{Gaia} DR3\\
       & Astrometry + RV & RV-only &  Astrometry-only \\
       \hline
        $a$ & 0.22$\pm$0.08 & 0.208317$\pm$0.000020 &  0.188$\pm$0.04 \\
        $e$  & 0.10$\pm$0.02 &  0.0324$\pm$0.0013 & 0.16$\pm$0.15 \\
        $P$  & 61.02$\pm$0.03 & 61.1166$\pm$0.0086 & 61.4$\pm$0.2\\
        $i$ & 84$\pm$6  &  59 & 101$\pm$8 \\
        $\omega_p$ & 158.96$\pm$0.36 &  230.3$\pm$3.2 &  78$\pm$57\\ 
        $\Omega$ &  25$\pm$4 & -- &  4.2$\pm$4.5 \\
        $T_0$ & 2450107.9$\pm$1.9 & 2450546.8$\pm$0.5$^{*}$ &  2457381$\pm$18$^{\ddagger}$\\
        M$_{p}$& 600$\pm$108 & 723.2235$\pm$1.4302 & 1144$\pm$127  \\
        \hline
    \end{tabular}
    \caption{GJ 876 b System Parameters and Orbital Elements Used in Simulations \\
     NOTE -- $^{*}$\citealt{rivera2010} give their timing parameter as the mean anomaly from the reference date 2450602.093, with mean anomaly~=~295.0$\pm$1.0, from which we computed the tabulated $T_0$; $^{\ddagger}$The $T_0$ given in the NSS is 2015.98$\pm0.05$ yr, which converts to the tabulated JD value.
     }
    \label{tab:gj876borb}
\end{table}

\subsection{The Habitable-Zone Terrestrial Planet Proxima Centauri b}\label{sec:proxcen}

Proxima Centauri b (hereafter Prox Cen b) is a (likely) terrestrial planet in its star's habitable zone. It is among the top priority targets for the first direct detection of a potential Earth-like exoplanet. %The contrast of habitable zone planets around M dwarfs is favorable compared to their counterparts around more massive stars. 
Despite the small separation of the habitable zone around M dwarfs, Proxima Cen b’s close proximity to our solar system means its angular separation is potentially accessible for direct detection by upcoming ELTs with ExAO coronagraphic instruments. Detection of Prox Cen b in reflected light is the ultimate science goal of the current ExAO instrument MagAO-X on the Magellan Clay Telescope \citep{Males2022MagAOXSPIE}.

Prox Cen b was discovered with RV using the HARPS spectrograph \citep{AngaldaEscude2016}. Recent efforts by \citet{Suarez2020} and \citet{Faria2022ProxCen} using the ESPRESSO instrument for RV monitoring have refined orbital elements for Prox Cen b, as well as the discovery of Prox Cen d with a $\approx 5$ day period. Table \ref{tab:ProxCenb-orbital-elements} shows orbital element fit results for Prox Cen b from the ESPRESSO RV data added to past data. 

\begin{figure*}
    \centering
    \includegraphics[width=0.48\linewidth]{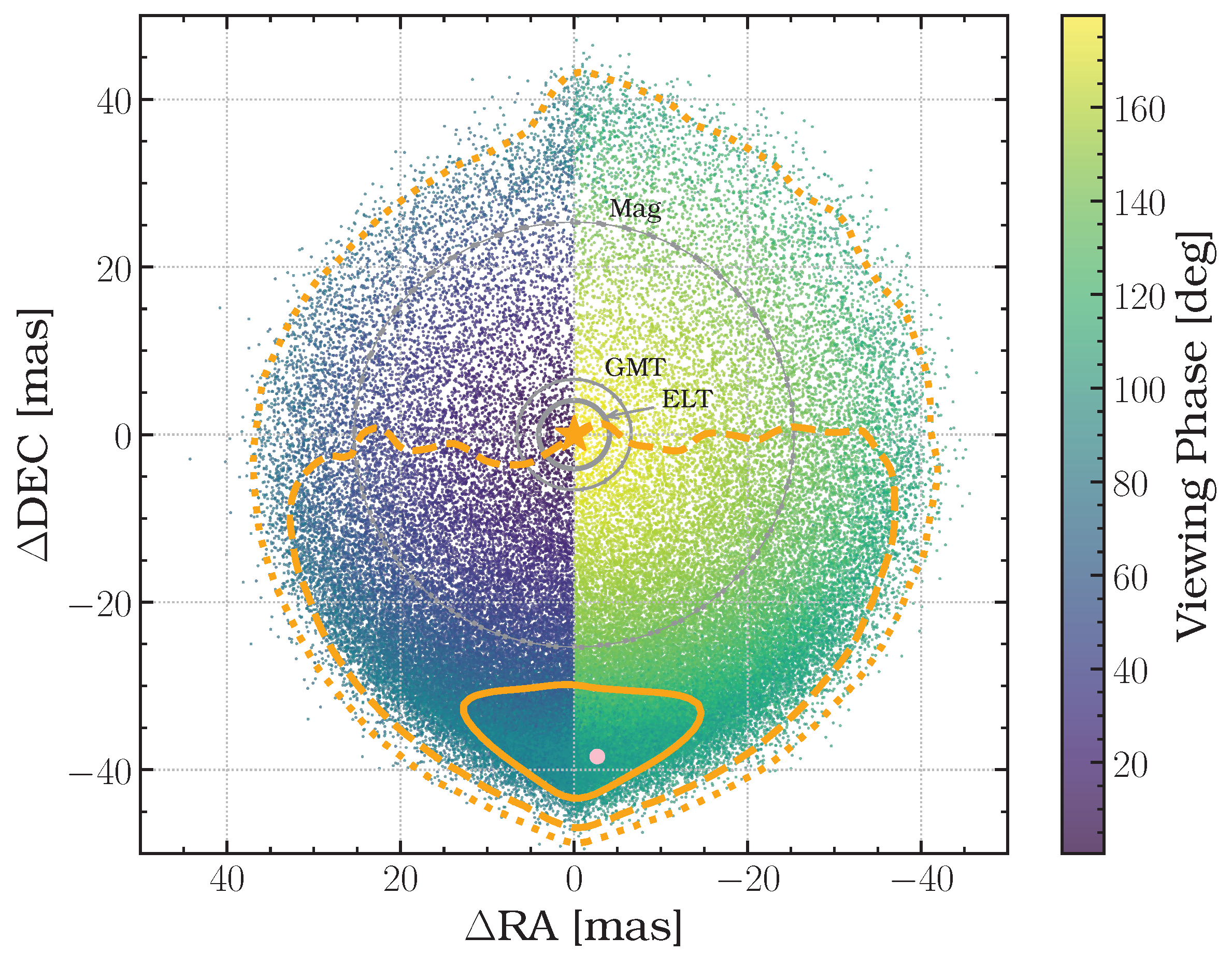}
    \includegraphics[width=0.48\linewidth]{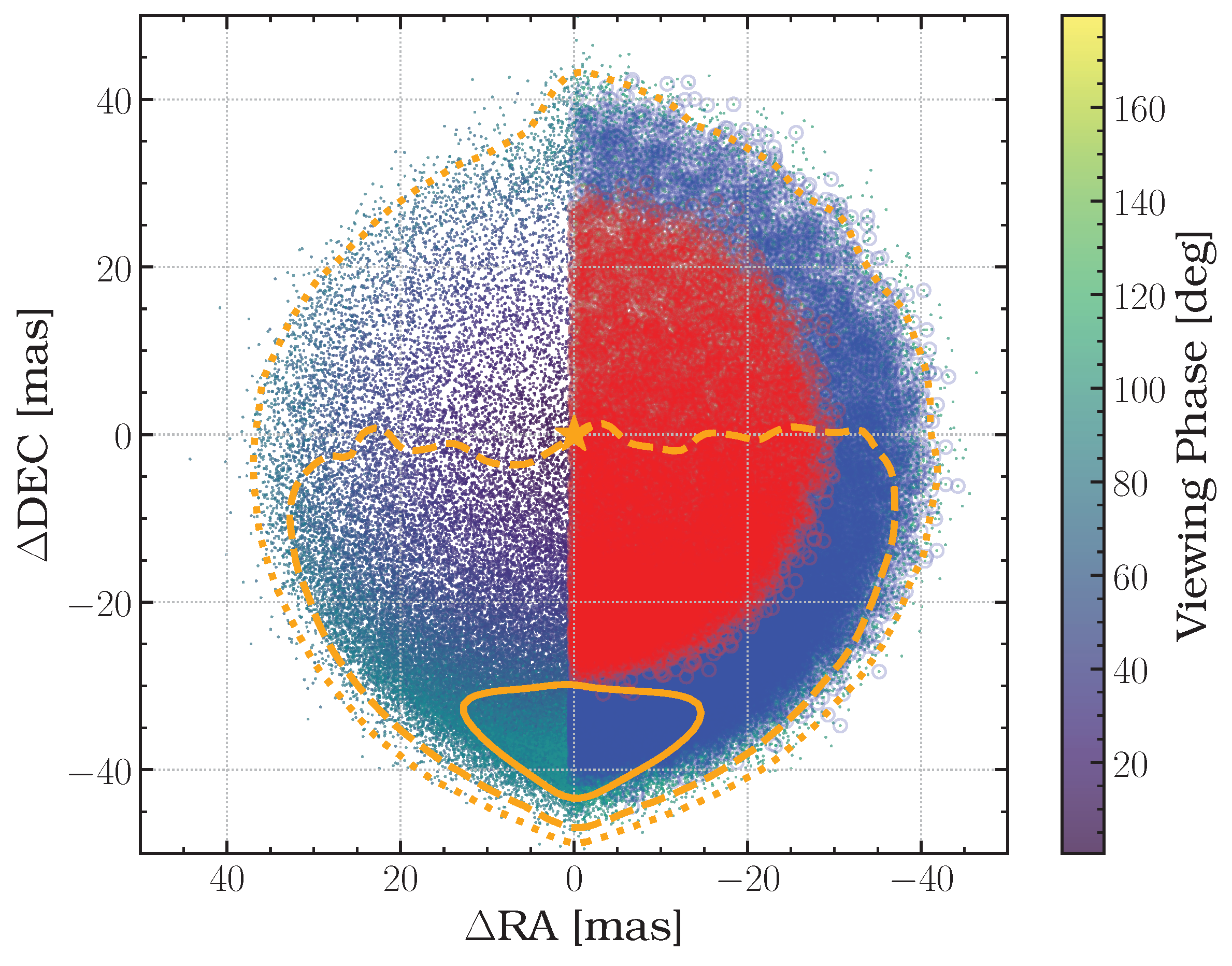}
    \caption{Two views of the predicted sky position of Prox Cen b at the expected time of maximum elongation from the star drawn from parameters of Table \ref{tab:ProxCenb-orbital-elements}. \textit{Left}: Scatter plot of all the simulated realizations of possible sky positions at the time Prox Cen b is predicted to be at it's maximum elongation from the star. The colormap shows the viewing phase at that point along the orbit; orange curves mark 1-, 2-, and 3$\sigma$ contours. The grey circles mark the size of 2$\lambda/D$ for ELT-sized primary (39~m, solid), GMT-sized primary (24.5~m, dotted), and Magellan Clay Telescope-sized primary (6.5~m, dash-dotted) at 0.8$\mu$m. The pink marker displays the expected location of the planet from the mean parameter values. \textit{Right}: same as left. Points at viewing phases where expected contrast drops by more than 1/2 compared to quadrature are blue, phases more than an order of magnitude fainter than quadrature are marked in red.
    }
    \label{fig:Proxcenb-clouds}
\end{figure*}

\begin{figure}
    \centering
    \includegraphics[width=1\linewidth]{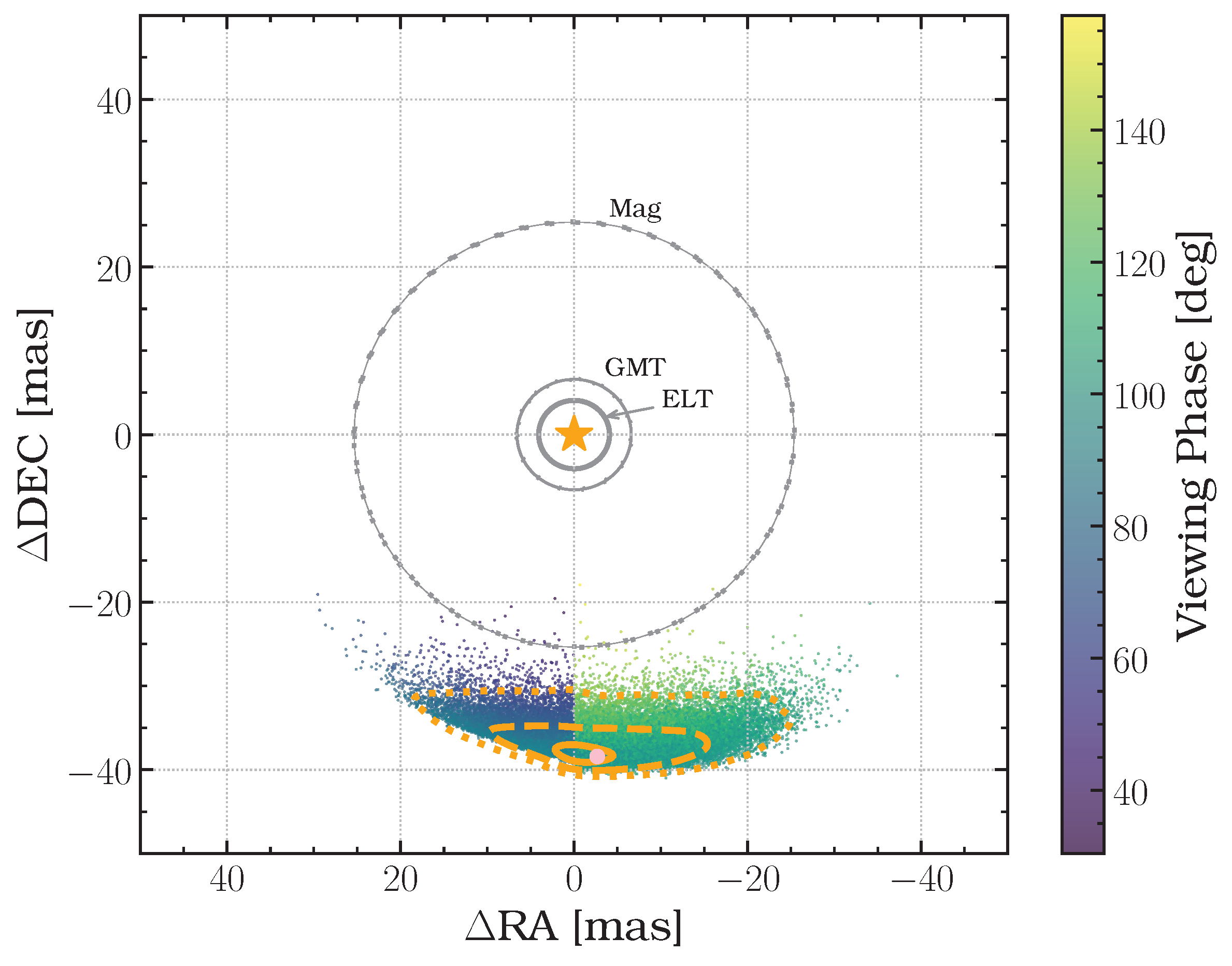}
    \caption{Same as Figure \ref{fig:Proxcenb-clouds} but with uncertainty on $\sigma_{T_0} = 10$~hours and $\sigma_{\omega_p} = 2^\circ$.
    }
    \label{fig:Proxcenb-clouds2}
\end{figure}

%We used the \citet{Suarez2020} orbital solution and \texttt{projecc} to predict the planet's location for direct imaging. 
\citet{Faria2022ProxCen} has the most recent orbit solution, from fit to ESPRESSO + literature data. They define the timing parameter as mean anomaly past the first observation date, BJD = 2458524.857066. The Exoplanet Archive lists $\omega_*$ in degrees however the \citet{Faria2022ProxCen} orbit was fit with $\omega$ in radians, and the reported result in the Exoplanet Archive is thus in radians, not degrees as indicated in the Archive entry. They use both cross-correlation function and template matching (TM) algorithms for RV extraction, and find that TM is more precise, so we adopt those values, given in Table \ref{tab:ProxCenb-orbital-elements}. \citet{Suarez2020} actually has tighter constraints on $T_0$ and $\omega$ (Table 3) than the more recent \citet{Faria2022ProxCen} solution, so we used that solution for analysis and Figures \ref{fig:Proxcenb-clouds} and \ref{fig:Proxcenb-clouds2}.

\begin{table}[]
    \centering
    \begin{tabular}{ccccc}
       Parameter & \citet{Faria2022ProxCen} & \citet{Suarez2020}  \\
       \hline
        $a$ & $0.04856\pm0.00030$ & $0.04856\pm0.00029$ \\
        $e$ & $0.02^{+0.04}_{-0.02}$ & $0.109^{+0.076}_{-0.68}$ \\
        $P$ &  $11.18688^{+0.0029}_{-0.0031}$ & $11.18418^{+0.00068}_{-0.00074}$ \\
         $\omega_p^*$ & $9\pm114$ & 71$\pm$42 \\ 
         $M^{**}$ & 5.0$^{+1.9}_{-2.1}$ & -- \\
        $T_0$ - 2450000 &  $8530.2\pm1.3$  & 8530$^{+1.3}_{-1.4}$ \\
        M$_{*}$  & $0.1221\pm 0.0022$ & -- \\
        M$_{p} \sin(i)$ &  $1.07\pm0.11$ &1.16$\pm0.13$ \\
        $\varpi$ &$768.50\pm0.20$ & -- \\
        \hline
    \end{tabular}
    \caption{Prox Cen b System Parameters and Orbital Elements Used in Simulations \\
    $^{*}$\citet{Faria2022ProxCen} tabulates $\omega_*$ in radians in Table C.1 (which is propagated to the Exoplanet Archive), we have tabulated the resulting $\omega_p$ in degrees; $^{**}$\citet{Faria2022ProxCen} parameterizes the timing parameters as the mean anomaly $M$ since the first observation date, BJD = 2458524.857066; we tabulate here their published value of $M$ and the resulting $T_0$. \citet{Suarez2020} elements are taken from their Gaussian Process fit to the full dataset with 2 signals (Table 3). }
    \label{tab:ProxCenb-orbital-elements}
\end{table}

Despite recent ESPRESSO data, the epoch of periastron passage is not well constrained, with an uncertainty of $\sigma_{T_0}=1.3$ days, more than a 1/10th of 11.2 day period. This, and the large uncertainty on the related parameter $\omega_p$, makes it challenging to predict the optimal observing time for catching the planet at maximum elongation (when detection is most favorable). Figure \ref{fig:Proxcenb-clouds} (left) shows a cloud of randomly selected possible planet locations drawn from the priors given in Table \ref{tab:ProxCenb-orbital-elements} at a time it is expected to be at max elongation (2025-04-16 05:57:22 UTC) and $\Omega$ artificially fixed at $\Omega = 0^{\circ}$ for clarity.  The planet phase is shown by the colormap. The grey circles mark $\lambda/D$ for an ELT (solid grey line) and GMT-sized primary (thin grey line) at visible wavelengths\footnote{We adopt 1$\lambda/D$ s the optimistic coronagraph performance envelope to illustrate the detection problem \citep{Males2024GMagAOX}}. As Prox Cen b detection is the ultimate science goal of MagAO-X, we also show 1$\lambda/D$ for a Magellan-sized (6.5~m) primary (grey dotted line). We see that if we try to observe Prox Cen b at the optimal time (max elongation, marked by the pink marker), there is a 7\% chance it will be closer that 4$\lambda/D$ for ELT (a typical separation for highest contrast with a coronagraph), a 25\% chance it will be closer than 4$\lambda/D$ for the GMT
%1\% chance it will actually be closer than 2$\lambda/D$ for ELT, a 4\% chance it will actually be closer than 2$\lambda/D$ for GMT
, and a 19\% chance it will be closer than $1\lambda/D$ for the Magellan Clay Telescope. Additionally, at the max elongation point we expect Prox Cen b to be at quadrature ($\alpha = 90^\circ$), however 34\% of simulated realizations are at $\alpha > 110^\circ$, where contrast drops by a factor of 2 (Figure \ref{fig:Proxcenb-clouds}, right, blue circles), and 11\% are at phases with contrast more than an order of magnitude lower than the contrast expected at quadrature ($\alpha > 137^\circ$; Figure \ref{fig:Proxcenb-clouds}, right, red circles). With this level of uncertainty, there is a significant chance that the planet may be in a position where detection is impossible when it is expected to be at the optimal position. Thus, a non-detection would be uninformative as it would be unclear whether the planet is at the expected location but has different physical characteristics than predicted (e.g., a lower albedo) placing it below detection limits, or if it is instead at an unexpected location (e.g., much closer than anticipated), unknowingly placing it inside the inner working angle of the coronagraph or at a larger phase angle (higher contrast) than expected.

If inclination is constrained, the uncertainty in planet position is driven almost entirely by large uncertainties in $T_0$ and $\omega_p$. Figure \ref{fig:Proxcenb-clouds2} shows the same set of simulated points for Prox Cen b but with the uncertainty on $T_0$ artificially lowered to $\sigma_{T0} = $~10~hours and uncertainty on $\omega_p$ set to $\sigma_{\omega_p}=$~2$^\circ$. With those two parameters much better constrained the sky position is concentrated at the expected optimal observing point. Decreasing $\sigma_{T0}$ and $\sigma_{\omega_p}$ should be one of the top priorities for planning future reflected light detection observations for Prox Cen b. While this level of precision may be ambitious, it is representative of the kind of precision needed for optimum efficiency in observing this planet.

\section{Discussion}\label{sec:discussion}

\subsection{Challenges \& needed improvements to implement literature orbit solutions to predict planet locations}

During this exploration, we encountered difficulties in applying literature orbit solutions to predict planet locations.%, underscoring the need for improvements in enabling direct imaging of exoplanets discovered via astrometry and RV. 
 Beyond issues of differing and unstated parameter definitions and coordinate systems discussed above, we found that publications often omitted key orbital parameters necessary for prediction. Six parameters define an orbit in space, plus an additional timing parameter, such as the date of periastron passage or orbit fraction past a reference date.%, to describe the planet's position along the orbit as a function of time. 
 This timing parameter was frequently difficult to extract or entirely absent. %It is important to review the primary sources for orbital parameters to ensure that all of the relevant assumptions are accounted for in any analysis

%Additionally, using orbital parameter solutions straight from the Exoplanet Archive can introduce errors. We encountered one case where a published solution with angular units in radians was misentered as degrees. In another case, the timing parameter in the paper was given as the mean anomaly of the periastron past a reference date, for which only the reference date was imported into the Archive as the epoch of periastron passage. The \citet{Benedict2002GJ876bMass} RV+astrometric solution for GJ~876~b is not given at all in the Exoplanet Archive. These issues highlight the need for caution when applying Exoplanet Archive parameters, particularly when analyzing large planetary populations, as individual orbit solution details may be misrepresented.\footnote{The authors wish to note the difficult and important work of the Exoplanet Archive and that this is not an implied criticism of this invaluable tool.}

Typically parameter covariances and the shape of non-Gaussian posteriors aren't reported in published orbit solutions. As a result, treating parameters as Gaussian and independent, as we have done here, can introduce errors when predicting the planet's location. For teams designing a direct imaging campaign for a planet with a marginally well-constrained position, where small differences in the error distributions are likely to matter, we recommend incorporating the full error information (covariances and asymmetric errors). This can be achieved either by performing independent orbit fitting using the RV and astrometric data to obtain the error distributions or collaborating closely with the teams that conducted the original fits. Accepting user-supplied covariances or full posteriors is planned for the next phase of \texttt{projecc} upgrades.

%\changed{While we explore in detail here only Prox~Cen~b and GJ~876~b, in the course of this analysis we examined literature orbit solutions for 19 potential reflected light targets: Gas giants HR 810 b, 70 Vir b, HD 128311 b, HD 114783 b; Potenitally terrestrial planets 61 Vi c, HD 192310 b, HD 102365 b, GJ 876 g, $\tau$ Cet g, h, e, and f, HD~180617~b, $\beta$~Pic~c; and potential \textsl{Roman} CGI targets $\epsilon$ Eri b, 47~UMa~c, and HD~219134~h \citep{Carrion-Gonzalez2021RomanReflLightTargets}. Published orbit solutions and \texttt{projecc} plots for each are available on the webapp, and the number of pre-loaded planet solutions will continue to grow and evolve. We found that most of the potentially rocky planets \todo{more}}

\subsection{Uncertainties in sky position are driven mainly by uncertainties in $\omega$ and $T_0$}

We examined the possible sky positions of Prox Cen b when the mean orbital solution predicts maximum elongation, the optimal time for observation. While many planets remain accessible to direct imaging over a larger fraction of their orbit, precise predictions are essential for planning. We used the maximum elongation time as a test case and found that well-determined sky positions require small uncertainties in the periastron position $\omega_p$ and time $T_0$.

\begin{figure}
    \centering    \includegraphics[width=1\linewidth]{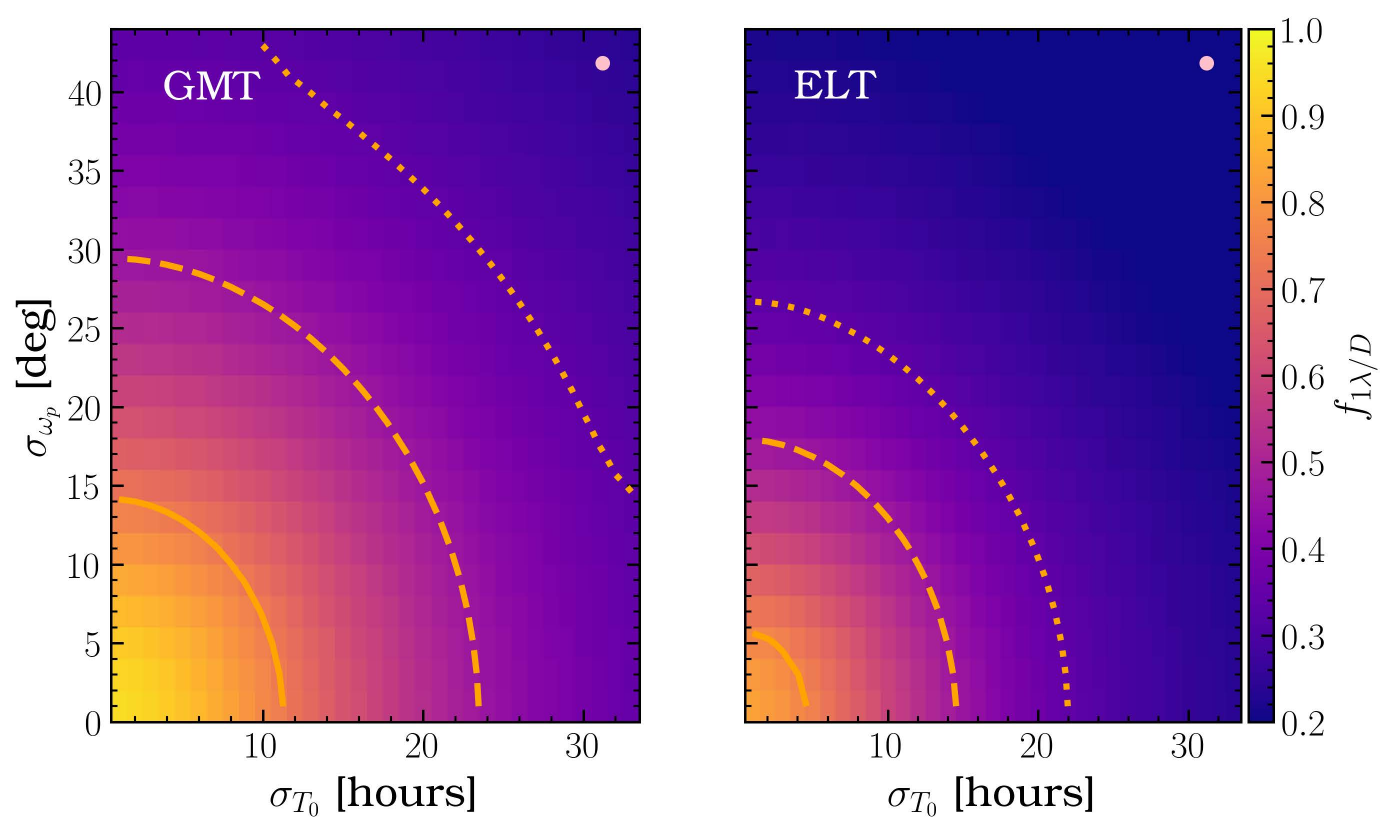}
    \caption{Quantifying improvement needed in $\omega_p$ and $T_0$ uncertainties for Prox Cen b. Left: fraction of orbit simulations within an aperture of radius = 1$\lambda/D$ for a GMT-sized primary at 0.8$\mu$m centered at the expected planet location for varying values of $\sigma_{\omega_p}$ and $\sigma_{T_0}$. Contours mark $f = 0.16, 0.5,$ and $0.84$. Right: same as left for an ELT-sized primary. The pink dot marks the current best values
    }
    \label{fig:sigmas}
\end{figure}

To quantify this, we performed a series of orbit simulations of Prox Cen b using the parameters listed in Table \ref{tab:ProxCenb-orbital-elements}, \citep{Suarez2020}, but with reduced uncertainties on $\omega_p$ and $T_0$ ($\sigma_{T_0} \in [1,34]$ hours in intervals of 1 hour; $\sigma_{\omega_p} \in [0^{\circ},44^{\circ}]$ in intervals of 1$^{\circ}$). We then placed an aperture of radius 1$\lambda/D$ at the max elongation location as predicted from the mean orbital parameters (pink dot in Figure \ref{fig:Proxcenb-clouds}, left) and determined the fraction of orbit simulation points that fell within the aperture.  Figure \ref{fig:sigmas} shows the result for a GMT-sized primary (left) and an ELT-sized primary (right), with the colorbar showing the fraction of points within the aperture and the contours marking $f=0.16, 0.5,$ and $0.84$. We see that to have a 50\% chance of finding the planet within 1$\lambda/D$ of expected location, we need $\sigma_{T_0} \lesssim 20$~hrs and $\sigma_{\omega_p} \lesssim 25^\circ$ for a GMT-sized primary; for the ELT this is even more strict at $\sigma_{T_0} \lesssim 15$~hrs and $\sigma_{\omega_p} \lesssim 15^\circ$ for this planet.

%In addition to the inconsistency noted above among $\omega$ vs $\omega_*$ vs $\omega_p$, many published RV solutions do not include a timing parameter, or are unclear how it was employed in their solution. $T_0$ and $\omega$ are dependent on each other to determine position at any given time, you cannot employ an $\omega$ value when predicting sky position without knowing the reference time it was determined from. There are various ways to report a timing parameter (e.g. epoch of periastron passage [as we have used], mean anomaly of periastron past a specific date [as in \citealt{rivera2010}], conjunction time [as in \citet{rosenthal2021}]), so being explicit about how it was determined is essential. 

\subsection{Precise orbit solutions don't guarantee accurate location predictions}

Giant planets such as GJ~876~b tend to have more precise orbital solutions than terrestrial planets due to larger RV and astrometric signals. However, in practice, even when the orbit appears well constrained, it does not always translate to an accurate prediction of the planet's location on the sky (e.g., Eps~Ind~Ab). To understand the causes of these discrepancies and improve our ability to predict planet positions for imaging based on RV and astrometry, it is crucial to continue efforts to directly image companions with well constrained indirect orbit solutions. %For example, directly imaging brown dwarfs and giant planets with seemingly well-constrained orbits would allow us to build up a population from which to test the accuracy of indirect orbit solutions and identify potential biases. 

If a planet's location is uncertain, a non-detection becomes ambiguous -- it may be due to the planet being fainter than expected based on its physical properties or simply due to an incorrect position and phase prediction. This results in a hit to observing efficiency, as many observations are required for the planet to be in a favorable position in at least some of them. As these telescopes will be in high demand, this drastically reduces the ability to conduct high-impact surveys with robust survey yields. We are planning a follow-on paper incorporating orbital uncertainties into reflected-light survey planning.

\subsection{Orbital Parameter Uncertainties Compound Over Time}

\changed{Currently \texttt{projecc} assumes published orbital parameter uncertainties remain valid at the proposed observation time. In reality, uncertainties increase with time since the last observation used in the fit \citep[e.g.][]{Deeg2015ParameterErrors, Mallonn2019Ephemeris}. }
\changed{To illustrate this effect, we used the \citet{rivera2010} RV-only fit for GJ~876~b, which has a relatively well-constrained solution. We ran a Monte Carlo simulation over 100 periods, from the last observation used in the fit (JD 2455202.74) to early 2027 (JD 2461406.5). Beginning at JD 2455202.74 we drew each orbital parameter from a normal distribution defined by the publish uncertainties, then propagated these uncertainties to each point in the array to generate a new array of values for the next period. We repeated this process for 100 periods, using each preceding period's uncertainty to determine the range for the next. After about 25 periods ($\approx$4 years) the semi-major axis mean increases exponentially; after about 10 periods eccentricity increases exponentially and is essentially unconstrained after about 25 periods; $\omega$ is essentially unconstrained after about 15 periods ($\approx$2.5 years); $\sigma_{t0}$ becomes larger than one period after about 14 periods. Thus, this solution is reliable for only a few years beyond the last constraining observation. For reflected-light campaigns, acquiring new measurements and updated fits close to the observing date will significantly improve location predictions. 
}

\subsection{Proxima Centauri b is the ``easiest" HZ terrestrial planet in reflected light}

Proxima Centauri b is the golden goose of exoplanet detection and characterization. Its direct detection is the ultimate science goal of the ExAO instrument MagAO-X \citep{Males2024MagAOXPhaseIICommisioningResults} and its follow-on for the GMT (GMagAO-X, \citealt{Males2024GMagAOX}), the ExAO high-resolution spectrograph RISTRETTO on VLT \citep{Lovis2024Ristretto, Blind2025RistrettoPIAAN}, the SPHERE+ \citep{Stadler2024UpgradingSPHERE} ExAO instrument at VLT, a pathfinder for ELT-PCS \citep{Kasper2021-ELTPCS}, and may be detectable with ELT/HARMONI \citep{vaughanMaskCanHARMONI2024}. As we've shown, it is also difficult to predict its location at any given time with published orbits, despite much dedicated RV survey time. Despite its proximity, Prox Cen is a cool M dwarf (T$_{\rm{eff}}=2900\pm100$~K, \citealt{Faria2022ProxCen}) and faint (m$_{V} = 11$), making robust orbit parameter determination of this small planet difficult. Yet it is by far the easiest known terrestrial planet to direct imaging. Other Earth-sized planets ($\lesssim 2$R$_\oplus$) in or near their star's HZ in Figure \ref{fig:targets} are closer and fainter than Prox Cen b; most are close than 2~$\lambda/D$ for a GMT-sized primary. For example, GJ~1061~d (contrast $\approx5\times10^{-5}$, sep $\approx3.5\lambda/D$) and Teegarden's~Star~c (contrast = $\approx8\times10^{-5}$, sep $\approx2\lambda/D$) both receive similar stellar instellation to Prox Cen b but are more difficult to detect. Refining Prox Cen b's orbital elements is essential for enabling efficient reflected light detection of a habitable zone planet.

\section{Conclusion}
We have developed a community tool, called \texttt{projecc}, to assist planning direct imaging observations for planets with published RV and/or astrometric orbital solutions. We used \texttt{projecc} to analyze current literature orbit solutions for GJ~876~b and Proxima Centauri b, and showed that orbital elements of Proxima Centauri b need further constraints to enable efficient direct imaging campaigns.

More dedicated surveys to precisely and accurately measure orbital parameters, especially $\omega_p$ and $T_0$, for ground- and space-based planets that are or will be ideal for future direct imaging campaigns is recommended. We have also provided the interactive GRIP webapp (\url{http://getagrip.space}, \changed{DOI: 10.5281/zenodo.15829820}) to assist the community in determining where to invest orbit monitoring in the run up to ELTs, \textsl{Roman}, and HWO.

In addition, we produced the following conclusions:

\begin{itemize}
    % \item As direct imaging of RV and astrometrically detected planets emerges and is expected to expand rapidly, we recommend RV, astrometry, and direct imaging communities adopt a standard set of orbital parameter definitions, coordinate system, and reference directions, applied consistently. Papers reporting orbital elements should explicitly state these definitions and include all parameters necessary to predict planet locations as a function of time. Collaboration across these communities is essential to establish norms and to work together to design coordinated ground- and space-based direct imaging campaigns.
    \item Publications reporting orbital parameters from fits should make the full posterior distribution of all parameters available digitally in some way. Additionally making the raw data public would allow others to fit the data and obtain their own posterior distributions.
    
    \item Refining of $\omega_p$ and $T_0$, and clearly defining the timing parameter, should be a priority of RV and RV+astrometry orbit solutions.
    \item Knowing a priori the longitude of nodes is beneficial for high contrast imaging survey efficiency, which can only be determined astrometrically or through direct detection. 
    \changed{\item \texttt{projecc} does not currently account for the compounding of orbital parameter uncertainty over time, which can compound rapidly; measurements and fits close to observation time improve efficiency.}
    \item Proxima Centauri b is the ``easiest" terrestrial planet in the HZ for direct imaging from the ground, yet its orbit remains insufficiently constrained for cued observations for efficient direct detection. Refining its orbit should be a high priority for teams preparing for this difficult and important direct detection and atmosphere characterization.
\end{itemize}

%\begin{acknowledgments}

The authors wish to thank Elisabeth Matthews for helpful conversations and the anonymous referee for helpful comments that improved the manuscript and analysis. Also we want to thank Matson Garza for generating the incredible name for the web app. Excellent work!

L.A.P.~acknowledges research support from the University of Michigan through the ELT Fellowship Program.  

This research has made use of the NASA Exoplanet Archive, which is operated by the California Institute of Technology, under contract with the National Aeronautics and Space Administration under the Exoplanet Exploration Program.

This work has made use of data from the European Space Agency (ESA) mission
{\it Gaia} (\url{https://www.cosmos.esa.int/gaia}), processed by the {\it Gaia}
Data Processing and Analysis Consortium (DPAC,
\url{https://www.cosmos.esa.int/web/gaia/dpac/consortium}). Funding for the DPAC
has been provided by national institutions, in particular the institutions
participating in the {\it Gaia} Multilateral Agreement.

\software{Numpy \citep{harris_array_2020}, Astropy \citep{astropy_collaboration_astropy_2018}, Matplotlib \citep{hunter_matplotlib_2007}, NSS Tools \citep{Halbwachs2023GaiaAstrBinary}, scipy \citep{virtanen_scipy_2020}}

\facility{Exoplanet Archive}

%\end{acknowledgments}

\appendix
\section{Installing \texttt{projecc}}\label{appendixA2}

We have provided a tool to facilitate community preparation for direct imaging campaigns of known indirectly detected planets using the \texttt{projecc} framework. \texttt{projecc} is installable via the Python Package Index (PyPI) at \url{https://pypi.org/project/projecc/} and via pip with the terminal command 
\begin{lstlisting}[language=Python]
pip install projecc
\end{lstlisting}
or by cloning the GitHub repository at \url{https://github.com/logan-pearce/projecc} \changed{and at DOI 10.5281/zenodo.15830010}

\section{Streamlit App}\label{appendixA}

Additionally we have provided an interactive web app to facilitate working with a target list of known indirectly detected planets for ground- and space-based platforms, and using \texttt{projecc} to examine orbit solutions, located at \url{https://reflected-light-planets.streamlit.app/} \changed{(DOI: 10.5281/zenodo.15829820)}

\subsection{Reflected Light Target List}

Figure \ref{fig:targets} shows hundreds of planets from the Exoplanet Archive with orbital parameters as a function of Lambertian contrast and star-planet separation. The web app provides the database of these targets along with a Jupyter notebook showing how the target list was compiled. The database is a list of RV-detected planets pulled from the Exoplanet Archive and filtered to provide the best set of planets for drawing a target list for reflected light imaging campaigns with ELTs and space-based platforms. This database is query-able with SQL. SQL is a language for selecting specific elements from a database. You can use the SQL interface above the database to query and filter the target list. After entering a query in the text box, the app will display the results and everything on the main page will update to use only the results from the query. To return to the whole database, reload the page. You can download the database or a query result using the tools in the upper right. Example queries are available on the readme page on the app.

The plot below the database shows the targets as a function of planet-star maximum separation in lambda/D vs uniform Lambertian planet-star flux contrast, sized by their estimated radius. By default it shows this for lambda = 0.8 um, diameter = 24.5 m (GMT sized primary), and geometric albedo = 0.3. Sliders below the plot allow you to adjust these parameters. For example, to see the planets as they would appear for a 6.5 m telescope (such as Magellan Clay or JWST), slide the diameter slider to 6.5. The plot will automatically update to the results of the SQL query, so planets you select in the query will be displayed in the plot. The plot is interactive based on Bokeh, so you can zoom and pan using the tools in the upper right corner, and hover over points to see details of each planet. You can also supply an estimated contrast curve to the plot. Click the "Add Contrast Curve" button, enter a list of separations in mas and corresponding contrast limits in flux units and hit enter.  The plot will update to show the curve.

\subsection{Using projecc to predict planet locations}
The "Predict Planet Location" tab has a user interface for working with the `projecc` package to generate prediction of a planet's location as a function of time from literature orbital parameter distributions.  

There are two ways to use the app interface.

\textit{\textbf{Manual parameter entry}}. Supply Gaussian mean, std values separated by a comma for each parameter in the units indicated to the form on the left side of the interface. By default $i$ is set to nan and $\Omega$ to zero. Indicate if the supplied planet mass represents a true mass or an M*sin(i) value. Supply a desired observation date, or alternately you can use the date at which the planet is expected to be at maximum separation from the star. If you elect to plot an aperture, \texttt{projecc} will compute the fraction of simulated realizations that fall within and aperture of radius 1$\lambda/D$ from the expected position and display this value below the returned figure. Once submitted, \texttt{projecc} will compute 100,000 simulated realizations of the planet's position on that date and display a plot on the right hand side, color coded by viewing phase. The fraction of points within an aperture and the spread of the points, as well as the entered parameter values, are displayed below the plot, along with a button for saving this figure if desired.

\textit{\textbf{Supplied planets}}. We also include all the planets and literature solutions examined in the course of preparing this work: 5 gas giants, 7 smaller planets, and the three most accessible \textsl{Roman} CGI planets from \citet{Carrion-Gonzalez2021RomanReflLightTargets} as of this writing, with more planning to be added. Planets in the target list plot that are highlighted in orange have literature solutions available to plot. Select a planet from the dropdown menu, then select a published set of orbital elements. If the published orbit solution doesn't constrain longitude of nodes, a radio button will appear to select if $\Omega$ will be drawn from U[0,360] or to fix it to zero. Then you can select to plot an aperture and the observation date. Once submitted, \texttt{projecc} will compute 100,000 simulated realizations of the planet's position on that date and display a plot on the right hand side, color coded by viewing phase. The fraction of points within an aperture and the spread of the points, as well as the orbital parameter values, are displayed below the plot, along with a button for saving this figure if desired.

More details and examples are available on the Read Me page of the app. We provide two summary statistics to quantify the confidence in locating a planet with a given literature solution. If selected, \texttt{projecc} will place an aperture at the expected location on the date specified, and display the fraction of simulated points within the aperture; a higher fraction is a more tightly constrained prediction. We also quantify the standard deviation of separations of simulated points and normalize the standard deviation by the expected separation; a lower number indicates points are more closely constrained to the expected separation.

Readers interested in collaboration or contribution are invited to reach out to the corresponding author.

\section{\changed{Derivation of ELT reflected light targets from the Exoplanet Archive}}\label{appendixReflLightTargets}

\changed{Due to the difficulty of reflected light planet observations, targeting known planets, with some parameters constrained or estimated such as mass, radius, and viewing phase, will be advantageous over blind searches.} Figure \ref{fig:targets} shows 417 of the nearest ($<$70~pc) RV detected exoplanets in the Exoplanet Archive\footnote{\url{https://exoplanetarchive.ipac.caltech.edu/}}. \changed{We queried the Archive on on 2025 Feb 25 for planets discovered via radial velocity with system distance $<70$~pc using the SQL string}
\begin{lstlisting}[
           language=SQL,
           showspaces=false,
           basicstyle=\ttfamily,
           numbers=left,
           numberstyle=\tiny,
           commentstyle=\color{gray}
        ]
SELECT * FROM pscomppars WHERE sy_dist < 70 AND discoverymethod = `Radial Velocity'
\end{lstlisting}
\changed{The distance limit of 70~pc was arbitrarily chosen to deliver $<1000$ results. This query returned 789 results. We removed entries without $a$, $P$, $e$, $\omega$, and $M_*$, leaving 704 entries. If no $i$ was provided we assigned inclination = 60$^{\circ}$, the average inclination for a uniform half-sphere; if mass was not available we used M$_p\sin(i) / \sin(i)$. If planet radius was not available we used an empirical mass/radius relationship (Appendix \ref{appendixD}). If stellar \Teff, radius, or spectral type were unavailable we inferred them from the relations of \citet{pecaut_intrinsic_2013}\footnote{Accessed from \url{https://www.pas.rochester.edu/~emamajek/EEM\_dwarf\_UBVIJHK\_colors\_Teff.txt}}. }

Figure \ref{fig:targets} shows a subset of this list with maximum projected separation $<20 \lambda/D$ for the GMT, $<32 \lambda/D$ for the ELT. The x-axis is the maximum on-sky separation in visible wavelengths (0.8~$\mu$m) for a 25.4~m primary mirror (the GMT primary, bottom) and a 39~m primary mirror (the ELT primary, top). The y-axis is the Lambertian flux contrast for a uniform sphere with geometric albedo A$_g = 0.45$ (arbitrarily chosen between the approximate albedo of Earth, $\approx0.4$, and Jupiter, $\approx0.5$), and the colormap gives the viewing phase angle at the maximum separation. We computed the planet-star flux contrast for a uniform reflecting Lambertian sphere as:
\begin{equation}\label{eqn: cahoy}
    C(\alpha,\lambda) = A_g \left(\frac{R_p}{r}\right)^2 \left[\frac{\sin\alpha + (\pi - \alpha)\cos\alpha}{\pi} \right]
\end{equation}
where $C(\alpha,\lambda)$ is planet-star contrast, $ A_g$ is geometric albedo, $R_p$ is planet radius, $r$ is planet-star true separation in the orbit plane, and $\alpha$ is defined in Eqn \ref{eqn:alpha} \citep[][]{cahoyExoplanetAlbedoSpectra2010, batalha_exoplanet_2019, lacyCharacterizationExoplanetAtmospheres2019}. We used a simple uniform geometric albedo model for this calculation and do not attempt to model potential atmospheres, which is beyond the scope of this analysis.\changed{ A notebook outlining the construction of this list, as well as an interactive version of Figure \ref{fig:targets}, are available on the GRIP webapp (\url{http://getagrip.space})}.

\section{Empirical Mass-Radius Relation for Lightly Irradiated Planets}\label{appendixD}

By lightly irradiated we mean (roughly) planets at Mercury's separation or further, scaled for stellar luminosity by $\sqrt{L_*}$. Here we make use of the transition from rocky to gas-dominated composition at $ 1.6  R_\oplus $ identified by \citet{rogers_2015} \citep[see also][]{marcy_2014}.  Below this radius we assume Earth composition and so $R \propto M^{1/3}$.  Above this we scale with a power law matched to the mean radius and mass of Uranus and Neptune, which defines the radius between $ 1.6^3 M_\oplus $ and this Uranus/Neptune mean point.  Above this point we use a polynomial fit in log(M) to points including the Uranus/Neptune mean, Saturn, Jupiter, and above Jupiter's mass the average points from the 4.5 Gyr 1 AU models from
\citet{fortney_2007}.  Above 3591.1 $ M_\oplus $ ($\approx 11 M_{jup}$) we scale as $M^{-1/8} $ based on the curve shown in \citet{fortney_2010}.
  
  \begin{equation}
    \frac{R}{R_\oplus} =
    \begin{cases}
      \left(\frac{M}{M_\oplus}\right)^{1/3}, & M < 4.1 M_\oplus \\
      0.62\left(\frac{M}{M_\oplus}\right)^{0.67}, & 4.1 M_\oplus \le M < 15.84 M_\oplus \\
      14.0211 - 44.8414 \log_{10}\left(\frac{M}{M_\oplus}\right) + \\ \qquad 53.6554 \log_{10}^2\left(\frac{M}{M_\oplus}\right)
        -  25.3289\log_{10}^3\left(\frac{M}{M_\oplus}\right) +  \\ \qquad 5.4920\log_{10}^4\left(\frac{M}{M_\oplus}\right) -  0.4586
  \log_{10}^5\left(\frac{M}{M_\oplus}\right), & 15.84 \le M < 3591.1 M_\oplus \\ 32.03
  \left(\frac{M}{M_\oplus}\right)^{-1/8}, & 3591.1 M_\oplus \le M 
  \end{cases} 
\label{eqn:adhoc-m2r}
\end{equation}

\begin{figure}
\centering
\includegraphics[width=6in]{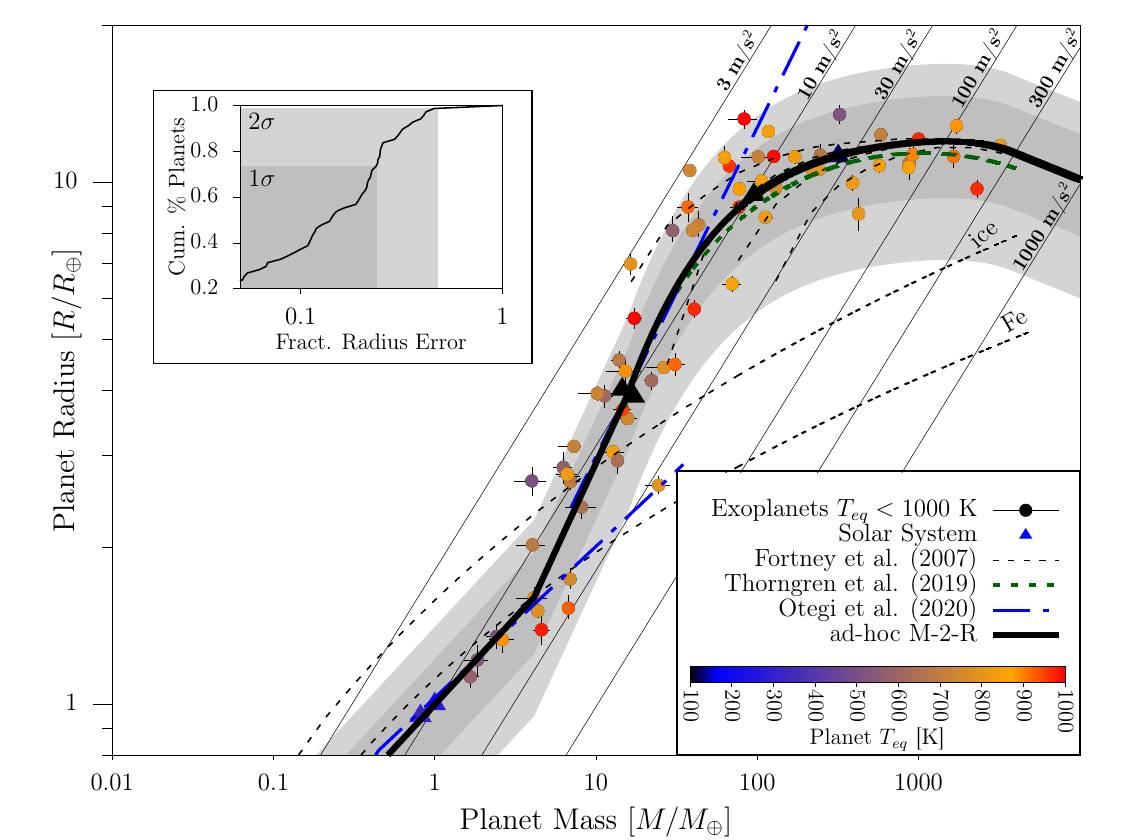}
\caption{Comparison of the empirical mass-radius relationship of Eqn. (\ref{eqn:adhoc-m2r}) to known planets with measured $M$ and $R$, and $T_{eq} < 1000$ K.  Given $M$, $R$ can be constrained to  $24\%$ at $1\sigma$, and $48\%$ at $2\sigma$. Also shown are the empirical and theoretical mass-radius curves of \citet{fortney_2007}, \citet{2020A&A...634A..43O}, and \citet{2019RNAAS...3..128T} as well as surface gravity contours.\label{fig:mass_radius}}
\end{figure}

Fig. \ref{fig:mass_radius} compares the predictions for this empirical mass-radius relation with planets in the NASA Exoplanet Archive \citep{2013PASP..125..989A} with both $M$ and $R$ measurements.  Missing stellar parameters were interpolated from the table of \citet{pecaut_intrinsic_2013}. A Bond albedo of 0.3 was assumed to calculate $T_{eq}$ and only planets cooler than 1000 K were kept to avoid thermal inflation \citep{2011ApJS..197...12D}.  These planets are plotted along with the Solar System planets. At low masses, Eq. \ref{eqn:adhoc-m2r}  matches well the empirical relationship of \citet{2020A&A...634A..43O}, and of \citet{2019RNAAS...3..128T} at higher masses.  The inset shows the distribution of errors for the shown planets, giving $1\sigma=24\%$, $2\sigma=48\%$.  The shaded regions around the black curve correspond to these uncertainties.  

\bibliography{references}{}
\bibliographystyle{aasjournal}

\end{document}